\documentclass[italian,english]{article}
\usepackage[T1]{fontenc}
\usepackage[latin1]{inputenc}
\usepackage{float}
\usepackage{color}
\usepackage{graphicx}
\usepackage{amssymb}

\makeatletter

 \newcommand{\lyxaddress}[1]{
   \par {\raggedright #1 
   \vspace{1.4em}
   \noindent\par}
 }

\usepackage{babel}
\makeatother
\begin{document}

\title{\textbf{Tuning the stochastic background of gravitational waves with
theory and observations}}

\author{\textbf{Christian Corda\ensuremath{¹}, Salvatore Capozziello\ensuremath{²},
Mariafelicia De Laurentis\ensuremath{³}}}

\maketitle

\lyxaddress{\begin{center}\ensuremath{¹}INFN - Sezione di Pisa and Università
di Pisa, Via F. Buonarroti 2, I - 56127 PISA\end{center}}

\lyxaddress{\begin{center}\ensuremath{²}Dipartimento di Scienze Fisiche, Università
di Napoli {}``Federico II'', INFN Sezione di Napoli, Compl. Univ
di Monte S. Angelo, Edificio G, via Cinthia, I-80126, Napoli, Italy\end{center}}

\begin{center}\ensuremath{³} Relativity and Gravitation Group Politecnico
di Torino, Corso Duca degli Abruzzi 24, I - 10129 Torino, Italy\end{center}

\lyxaddress{\begin{center}\textit{E-mail addresses:} \textcolor{blue}{\ensuremath{¹}christian.corda@ego-gw.it;
\ensuremath{²}capozzie@na.infn.it; \ensuremath{³} mariafelicia.delaurentist@polito.it}\end{center}}

\begin{abstract}
In this this paper the stochastic background of gravitational waves
(SBGWs) is analyzed with the auxilium of the WMAP data. We emphasize
that, in general, in previous works in the literature about the SBGWs,
old COBE data were used. After this, we want to face the problem of
how the SBGWs and f(R) gravity can be related, showing, vice versa,
that a revealed SBGWs could be a powerly probe for a given theory
of gravity. In this way, it will also be shown that the conformal
treatment of SBGWs can be used to parametrize in a natural way f(R)
theories.
\end{abstract}
\begin{itemize}
\item PACS numbers: 04.80.Nn, 04.30.Nk, 04.50.+h
\item Keywords: gravitational waves; extended theories of gravity; stochastic
background;
\end{itemize}

\section{Introduction}

The accelerate expansion of the Universe, which is today observed,
shows that cosmological dynamic is dominated by the so called Dark
Energy which gives a large negative pressure. This is the standard
picture, in which such new ingredient is considered as a source of
the \textit{rhs} of the field equations. It should be some form of
un-clustered non-zero vacuum energy which, together with the clustered
Dark Matter, drives the global dynamics. This is the so called {}``concordance
model'' (ACDM) which gives, in agreement with the CMBR, LSS and SNeIa
data, a good trapestry of the today observed Universe, but presents
several shortcomings as the well known {}``coincidence'' and {}``cosmological
constant'' problems \cite{key-1}. 

An alternative approach is changing the \textit{lhs} of the field
equations, seeing if observed cosmic dynamics can be achieved extending
general relativity \cite{key-2,key-3,key-4,key-5}. In this different
context, it is not required to find out candidates for Dark Energy
and Dark Matter, that, till now, have not been found, but only the
{}``observed'' ingredients, which are curvature and baryonic matter,
have to be taken into account. Considering this point of view, one
can think that gravity is not scale-invariant \cite{key-5} and a
room for alternative theories is present \cite{key-6,key-7,key-8}.
In principle, the most popular Dark Energy and Dark Matter models
can be achieved considering $f(R)$ theories of gravity \cite{key-2,key-3,key-4},
where $R$ is the Ricci curvature scalar. In this picture even the
sensitive detectors for gravitational waves, like bars and interferometers
(i.e. those which are currently in operation and the ones which are
in a phase of planning and proposal stages) \cite{key-9,key-10,key-11,key-12,key-13,key-14,key-15,key-16},
could, in principle, be important to confirm or ruling out the physical
consistency of general relativity or of any other theory of gravitation.
This is because, in the context of Extended Theories of Gravity, some
differences between General Relativity and the others theories can
be pointed out starting by the linearized theory of gravity \cite{key-6,key-17,key-18,key-19}. 

This philosophy can be taken into account also for the SBGWs which,
together with cosmic microwave background radiation (CMBR), would
carry, if detected, a huge amount of information on the early stages
of the Universe evolution \cite{key-2,key-20,key-21}). Also in this
case, a key role for the production and the detection of this graviton
background is played by the adopted theory of gravity \cite{key-21}. 

In the second section of this paper the SBGWs is analyzed with the
auxilium of the WMAP data \cite{key-20,key-22,key-23}. We emphasize
that, in general, in previous works in the literature about the SBGWs,
old COBE data were used (see \cite{key-24,key-25,key-26,key-27,key-28,key-29}
for example).

In the third section we want to face the problem of how the SBGWs
and f(R) gravity can be related, showing, vice versa, that a revealed
SBGWs could be a powerly probe for a given theory of gravity. In this
way, it will also be shown that the conformal treatment of SBGWs can
be used to parametrize in a natural way f(R) theories.

\section{Tuning the stochastic background of gravitational waves using the
WMAP data}

From our analysis, it will result that the WMAP bounds on the energy
spectrum and on the characteristic amplitude of the SBGWs are greater
than the COBE ones, but they are also far below frequencies of the
earth-based antennas band. At the end of this section a lower bound
for the integration time of a potential detection with advanced LIGO
is released and compared with the previous one arising from the old
COBE data. Even if the new lower bound is minor than the previous
one, it results very long, thus for a possible detection we hope in
the LISA interferometer and in a further growth in the sensitivity
of advanced projects.

The strongest constraint on the spectrum of the relic SBGWs in the
frequency range of ground based antennas like bars and interferometers,
which is the range $10Hz\leq f\leq10^{4}Hz$, comes from the high
isotropy observed in the CMBR.

The fluctuation $\Delta T$ of the temperature of CBR from its mean
value $T_{0}=2.728$ K varies from point to point in the sky \cite{key-20,key-22,key-23},
and, since the sky can be considered the surface of a sphere, the
fitting of $\Delta T$ is performed in terms of a Laplace series of
spherical harmonics

\begin{equation}
\frac{\Delta T}{T_{0}}(\hat{\Omega})=\sum_{l=1}^{\infty}\sum_{m=-l}^{l}a_{lm}Y_{lm}(\hat{\Omega}),\label{eq: fluttuazioni CBR}\end{equation}

and the fluctations are assumed statistically independent ($<a_{lm}>=0$,
$<a_{lm}a_{l'm'}^{*}>=C_{l}\delta_{ll'}\delta_{mm'}$). In eq. (\ref{eq: fluttuazioni CBR})
$\hat{\Omega}$ denotes a point on the 2-sphere while the $a_{lm}$
are the multipole moments of the CMBR. For details about the definition
of statistically independent fluctations in the context of the temperature
fluctations of CMBR see \cite{key-24,key-25}.

The WMAP data \cite{key-22,key-23} permit a more precise determination
of the rms quadrupole moment of the fluctations than the COBE data

\begin{equation}
Q_{rms}\equiv T(\sum_{m=-2}^{2}\frac{|a_{2m}|^{2}}{4\pi})^{\frac{1}{2}}=8\pm2\mu K,\label{eq: Q rms}\end{equation}

while in the COBE data we had \cite{key-25,key-26,key-27}

\begin{equation}
Q_{rms}=14.3_{-3.3}^{+5.2}\mu K.\label{eq: Q rms COBE}\end{equation}
 A connection between the fluctuation of the temperature of the CMBR
and the SBGWs derives from the \textit{Sachs-Wolfe effect} \cite{key-20,key-30}.
Sachs and Wolfe \textit{}showed that variations in the density of
cosmological fluid and GWs perturbations result in the fluctuation
of the temperature of the CMBR, even if the surface of last scattering
had a perfectly uniform temperature \cite{key-30}. In particular
the fluctuation of the temperature (at the lowest order) in a particular
point of the space is

\begin{equation}
\frac{\Delta T}{T_{0}}(\hat{\Omega})=\frac{1}{2}\int_{nullgeodesic}d\lambda\frac{\partial}{\partial\eta}h_{rr}.\label{eq: null geodesic}\end{equation}

The integral in eq. (\ref{eq: null geodesic}) is taken over a path
of null geodesic which leaves the current spacetime point heading
off in the direction defined by the unit vector $\hat{\Omega}$ and
going back to the surface of last scattering of the CMBR.

Here $\lambda$ is a particular choice of the affine parameter along
the null geodesic. By using conformal coordinates, we have for the
metric perturbation

\begin{equation}
\delta g_{ab}=R^{2}(\eta)h_{ab},\label{eq: metric perturbation}\end{equation}

and $r$ in eq. (\ref{eq: null geodesic}) is a radial spatial coordinate
which goes outwards from the current spacetime point. The effect of
a long wavelenght GW is to shift the temperature distribution of CMBR
from perfect isotropy. Because the fluctations are very small ($<\Delta T/T_{0}>\leq5*10^{-5}$
\cite{key-22,key-23}), the perturbations caused by the relic SBGWs
cannot be too large. 

The WMAP results give rather tigh constraints on the spectrum of the
SBGWs at very long wavelenghts. In \cite{key-24,key-25} we find a
constraint on $\Omega_{gw}(f)$ \textit{}which derives from the COBE
observational limits, given by

\begin{equation}
\Omega_{gw}(f)h_{100}^{2}<7*10^{-11}(\frac{H_{0}}{f})^{2}\textrm{ for }H_{0}<f<30H_{0}.\label{eq: limite COBE}\end{equation}

Now the same constraint will be obtained from the WMAP data \cite{key-20}.
Because of its specific polarization properties, relic SBGWs should
generate particular polarization pattern of CMBR anisotropies, but
the detection of CMBR polarizations is not fulfiled today \cite{key-31}.
Thus an indirect method will be used. We know that relic GWs have
very long wavelenghts of Hubble radius size, so the CMBR power spectrum
from them should manifest fast decrease at smaller scales (hight multipole
moments). But we also know that scalar modes produce a rich CMBR power
spectrum at large multipole moments (series of acoustic peaks, ref.
\cite{key-22,key-23}). Then the properties of tensor modes of cosmological
perturbations of spacetime metric can be extract from observational
data using angular CMBR power spectrum combined with large scale structure
of the Universe. One can see (fig. 1 ) that in the range $2\leq l\leq30$
(the same used in \cite{key-25}, but with the old COBE data \cite{key-32})
scalar and tensor contributions are comparable. From \cite{key-22,key-23},
the WMAP data give for the tensor/scala ratio $r$ the constraint
$r<0.9$. \foreignlanguage{italian}{(}$r<0.5$ \foreignlanguage{italian}{in
the COBE data, ref.} \cite{key-32}\foreignlanguage{italian}{); Novosyadly
and Apunevych} obtained $\Omega_{scalar}(H_{0})<2.7*10^{-9}$ \cite{key-31}.
Thus, if one remembers that, at order of Hubble radius, the tensorial
spectral index is $-4\leq n_{t}\leq-2$ \cite{key-20}, it results

\begin{equation}
\Omega_{gw}(f)h_{100}^{2}<1.6*10^{-9}(\frac{H_{0}}{f})^{2}\textrm{ for }H_{0}<f<30H_{0},\label{eq: limite WMAP}\end{equation}

which is greater than the COBE data result of eq. (\ref{eq: limite COBE}).

We emphasize that the limit of eq. (\ref{eq: limite WMAP}) is not
a limit for any GWs, but only for relic ones of cosmological origin,
which were present at the era of the CMBR decoupling. Also, the same
limit only applies over very long wavelenghts (i.e. very low frequencies)
and it is far below frequencies of the Virgo - LIGO band.

\begin{figure}
\includegraphics[%
  scale=0.9]{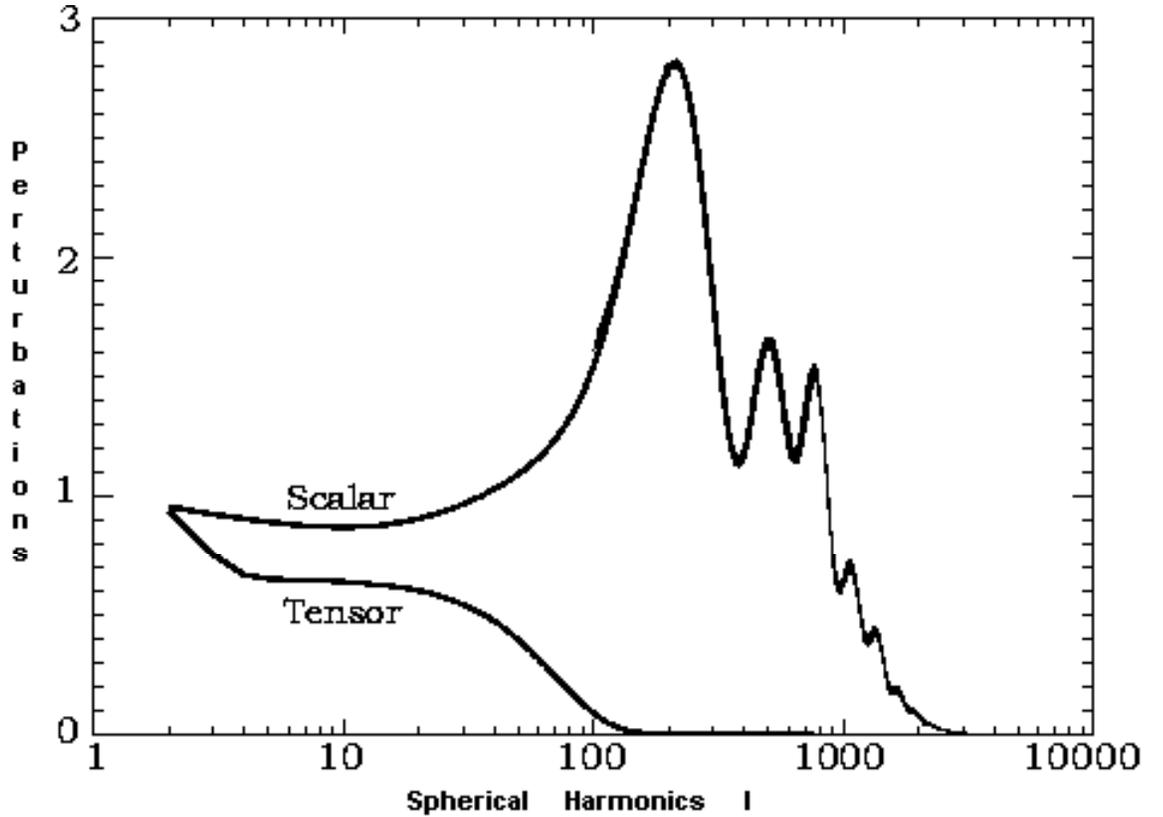}

\caption{The tensor to scalar ratio}
\end{figure}

The primordial production of the relic SBGWs has been analyzed in
\cite{key-25,key-26,key-27,key-33} (note: a generalization for f(R)
theories of gravity will be given in the next section following \cite{key-21}),
where it has been shown that in the range $10^{-15}Hz\leq f\leq10^{10}Hz$
the spectrum is flat and proportional to the ratio

\begin{equation}
\frac{\rho_{ds}}{\rho_{Planck}}\approx10^{-12}.\label{eq: rapporto densita' primordiali}\end{equation}

WMAP observations put strongly severe restrictions on the spectrum,
as we discussed above. In fig. 2 the spectrum $\Omega_{gw}$ is mapped:
the amplitude (determined by the ratio $\frac{\rho_{ds}}{\rho_{Planck}}$)
has been chosen to be \textit{as large as possible, consistent with
the WMAP constraint} (\ref{eq: limite WMAP}). \textit{}

\begin{figure}[H]
\includegraphics[%
  scale=0.9]{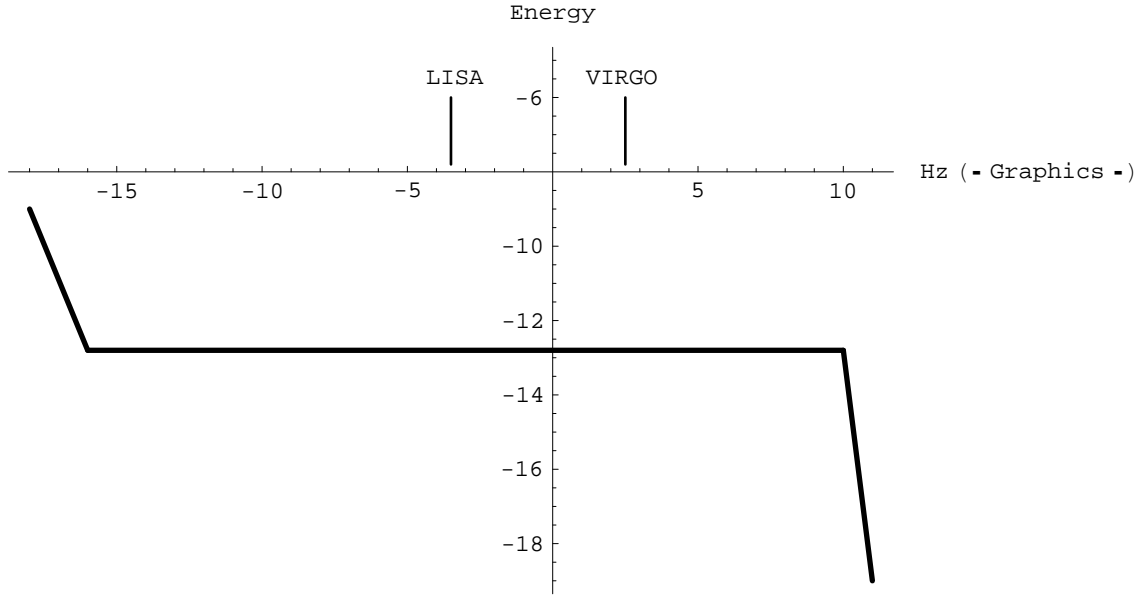}

\caption{The spectrum of relic gravitons in inflationary models is flat over
a wide range of frequencies. The horizontal axis is $\log_{10}$ of
frequency, in Hz. The vertical axis is $\log_{10}\Omega_{gr}$. The
inflationary spectrum rises quickly at low frequencies (wave which
rentered in the Hubble sphere after the Universe became matter dominated)
and falls off above the (appropriately redshifted) frequency scale
$f_{max}$ associated with the fastest characteristic time of the
phase transition at the end of inflation. The amplitude of the flat
region depends only on the energy density during the inflationary
stage; we have chosen the largest amplitude consistent with the WMAP
constrain discussed earlier: $\Omega_{gr}(f)h_{100}^{2}<1.6*10^{-9}$
at $10^{-18}Hz$. This means that at Virgo and Lisa frequencies, $\Omega_{gr}(f)h_{100}^{2}<9*10^{-13}$}
\end{figure}
Nevertheless, because the spectrum falls off $\propto f^{-2}$ at
low frequencies \cite{key-25,key-26,key-27,key-33}, this means that
today, at Virgo and LISA frequencies, indicated in fig. 2,

\begin{center}\begin{equation}
\Omega_{gw}(f)h_{100}^{2}<9*10^{-13},\label{eq: limite spettro WMAP}\end{equation}
\end{center}

while using the COBE data it was

\begin{center}$\Omega_{gw}(f)h_{100}^{2}<8*10^{-14}$(refs. \cite{key-25,key-32}).\end{center}

It is interesting to calculate the correspondent strain at $\approx100Hz$,
where interferometers like Virgo and LIGO have a maximum in sensitivity.
The well known equation for the characteristic amplitude \cite{key-20,key-24,key-25,key-29}
can be used:

\begin{equation}
h_{c}(f)\simeq1.26*10^{-18}(\frac{1Hz}{f})\sqrt{h_{100}^{2}\Omega_{gw}(f)},\label{eq: legame ampiezza-spettro}\end{equation}
obtaining

\begin{equation}
h_{c}(100Hz)<1.7*10^{-26}.\label{eq: limite per lo strain}\end{equation}

Then, because for ground-based interferometers a sensitivity of the
order of $10^{-22}$ is expected at $\approx100Hz$, four order of
magnitude have to be gained in the signal to noise ratio \cite{key-9,key-10,key-11,key-12,key-13,key-14,key-15,key-16}.
Let us analyze smaller frequencies too. The sensitivity of the Virgo
interferometer is of the order of $10^{-21}$ at $\approx10Hz$ \cite{key-9,key-10}
and in that case it is 

\begin{equation}
h_{c}(10Hz)<1.7*10^{-25}.\label{eq: limite per lo strain2}\end{equation}

For a better understanding of the difficulties on the detection of
the SBGWs a lower bound for the integration time of a potential detection
with advanced LIGO is released. For a cross-correlation between two
interferometers the signal to noise ratio (SNR) increases as 

\begin{equation}
(SNR)=\sqrt{2T}\frac{H_{0}^{2}}{5\pi^{2}}\sqrt{\int_{0}^{\infty}df\frac{\Omega_{gw}^{2}(f)\gamma^{2}(f)}{f^{6}P_{1}(|f|)P_{2}(|f|)}}.\label{eq: SNR2}\end{equation}

where $P_{i}(|f|)$ is the one-sided power spectral density of the
$i$ detector \cite{key-34} and $\gamma(f)$ the well known overlap-reduction
function \cite{key-34,key-35}. Assuming two coincident coaligned
detectors $(\gamma(f)=1)$ with a noise of order $10^{-48}/Hz$ (i.e.
a typical value for the advanced LIGO sensitivity \cite{key-36})
one gets $(SNR)\sim1$ for $\sim3*10^{5}years$ using our result $\Omega_{gw}(f)h_{100}^{2}\sim9*10^{-13}$
while it is $(SNR)\sim1$ for $\sim3*10^{7}years$ using previous
COBE result $\Omega_{gw}(f)h_{100}^{2}\sim8*10^{-14}$. Since the
overlap reduction function degrades the SNR, these results can be
considered a solid upper limit for the advanced LIGO configuration
for the two different values of the spectrum. 

The sensitivity of the LISA interferometer will be of the order of
$10^{-22}$ at $10^{-3}Hz$ \cite{key-37} and in that case it is 

\begin{equation}
h_{c}(10^{-3}Hz)<1.7*10^{-21}.\label{eq: limite per lo strain3}\end{equation}

Then a stochastic background of relic gravitational waves could be
in principle detected by the LISA interferometer. We also hope in
a further growth in the sensitivity of advanced projects.

We emphasize that the assumption that all the tensorial perturbation
in the Universe are due to a SBGWs is quit strong, but our results
(\ref{eq: limite spettro WMAP}), (\ref{eq: limite per lo strain}),
(\ref{eq: limite per lo strain2}) and (\ref{eq: limite per lo strain3})
can be considered like upper bounds. 

Reasuming, in this section the SBGWs has been analyzed with the auxilium
of the WMAP data, while previous works in literature, used the old
COBE data, seeing that the predicted signal for these relic GWs is
very weak. From our analysis it resulted that the WMAP bound on the
energy spectrum and on the characteristic amplitude of the SBGWs are
greater than the COBE ones, but they are also far below frequencies
of the earth-based antennas band. In fact the integration time of
a potential detection with advanced interferometers is very long,
thus, for a possible detection we have to hope in a further growth
in the sensitivity of advanced ground based projects and in the LISA
interferometer.

\section{Tuning the stochastic background of gravitational waves with f(R)
theories of gravity}

GWs are the perturbations $h_{\alpha\beta}$ of the metric $g_{\alpha\beta}$
which transform as three-tensors. Following \cite{key-21,key-38},
the GW-equations in the TT gauge are 

\begin{equation}
\square h_{i}^{j}=0,\label{eq: dalembert}\end{equation}
where $\square\equiv(-g)^{-1/2}\partial_{\alpha}(-g)^{1/2}g^{\alpha\beta}\partial_{\beta}$
is the usual d'Alembert operator and these equations are derived from
the Einstein field equations deduced from the Hilbert Lagrangian density
$L=R$ \cite{key-6,key-21}. Clearly, matter perturbations do not
appear in (\ref{eq: dalembert}) since scalar and tensor perturbations
do not couple with tensor perturbations in Einstein equations. The
Latin indices run from 1 to 3, the Greek ones from 0 to 3. Our task
is now to derive the analog of eqs. (\ref{eq: dalembert}) assuming
a generic theory of gravity given by the action \begin{equation}
A=\frac{1}{2k}\int d^{4}x\sqrt{-g}f(R),\label{eq: high order}\end{equation}

where, for a sake of simplicity, we have discarded matter contributions.
A conformal analysis will help in this goal. In fact, assuming the
conformal transformation \begin{equation}
\tilde{g}_{\alpha\beta}=e^{2\Phi}g_{\alpha\beta}\label{eq: conforme}\end{equation}

where the conformal rescaling \begin{equation}
e^{2\Phi}=f'(R)\label{eq: rescaling}\end{equation}

has been chosen being the prime the derivative with respect to the
Ricci curvature scalar and $\Phi$ the {}``conformal scalar field'',
we obtain the conformally equivalent Hilbert-Einstein action \begin{equation}
A=\frac{1}{2k}\int d^{4}x\sqrt{-\widetilde{g}}[\widetilde{R}+L(\Phi,\Phi_{;\alpha})],\label{eq: conform}\end{equation}

where $L(\Phi,\Phi_{;\alpha})$ is the conformal scalar field contribution
derived from

\begin{equation}
\tilde{R}_{\alpha\beta}=R_{\alpha\beta}+2(\Phi_{;\alpha}\Phi_{;\beta}-g_{\alpha\beta}\Phi_{;\delta}\Phi^{;\delta}-\frac{1}{2}g_{\alpha\beta}\Phi^{;\delta}{}_{;\delta})\label{eq: conformRicci}\end{equation}

and \begin{equation}
\tilde{R}=e^{-2\Phi}+(R-6[]\Phi-6\Phi_{;\delta}\Phi^{;\delta}).\label{eq: conformRicciScalar}\end{equation}

In any case, as we will see, the $L(\Phi,\Phi_{;\alpha})$-term does
not affect the GWs-tensor equations so it will not be considered any
longer (note: a scalar component in GWs is often considered \cite{key-17,key-39,key-40,key-41},
but here we are taking into account only the genuine tensor part of
stochastic background).

Starting from the action (\ref{eq: conform}) and deriving the Einstein-like
conformal equations, the GWs equations are \begin{equation}
\widetilde{\square}\widetilde{h}_{i}^{j}=0,\label{eq: dalembert conf}\end{equation}

expressed in the conformal metric $\tilde{g}_{\alpha\beta}.$ Since
no scalar perturbation couples to the tensor part of gravitational
waves it is 

\begin{equation}
\widetilde{h}_{i}^{j}=\widetilde{g}^{lj}\delta\widetilde{g}_{il}=e^{-2\Phi}g^{lj}e^{2\Phi}\delta g_{il}=h_{i}^{j},\label{eq: confinvariant}\end{equation}

which means that $h_{i}^{j}$ is a conformal invariant.

As a consequence, the plane wave amplitude $h_{i}^{j}=h(t)e_{i}^{j}\exp(ik_{i}x^{j}),$
where $e_{i}^{j}$ is the polarization tensor, are the same in both
metrics. In any case the d'Alembert operator transforms as 

\begin{equation}
\widetilde{\square}=e^{-2\Phi}(\square+2\Phi^{;\alpha}\partial_{;\alpha})\label{eq: quadratello}\end{equation}

and this means that the background is changing while the tensor wave
amplitude is fixed. 

In order to study the cosmological stochastic background, the operator
(\ref{eq: quadratello}) can be specified for a Friedman-Robertson-Walker
metric \cite{key-26,key-27}, and the equation (\ref{eq: dalembert conf})
becomes

\begin{equation}
\ddot{h}+(3H+2\frac{d\Phi}{dt})\frac{dh}{dt}+k^{2}a^{-2}h=0,\label{eq: evoluzione h}\end{equation}

being $\square=\frac{\partial}{\partial t^{2}}+3H\frac{\partial}{\partial t}$,
$a(t)$ the scale factor and $k$ the wave number. It has to be emphasized
that equation (\ref{eq: evoluzione h}) applies to any f(R) theory
whose conformal transformation can be defined as $e^{2\Phi}=f'(R).$
The solution, i.e. the GW amplitude, depends on the specific cosmological
background (i.e. $a(t)$) and the secific theory of gravity (i.e.
$\Phi(t)$). For example, assuming power law behaviors for $a(t)$
and $\Phi(t)=\frac{1}{2}\ln f'(R(t)),$ that is

\begin{equation}
\Phi(t)=f'(R)=f_{0}'(\frac{t}{t_{0}})^{m},\textrm{ $ $$ $$ $$ $$ $$ $$ $$ $ }a(t)=a_{0}(\frac{t}{t_{0}})^{n},\label{eq: phi a}\end{equation}

it is easy to show that general relativity is recovered for $m=0$
while 

\begin{equation}
n=\frac{m^{2}+m-2}{m+2}\label{eq: m n}\end{equation}

is the relation between the parameters for a generic $f(R)=f_{0}R^{s}$
where $s=1-\frac{m}{2}$ with $s\models1$ \cite{key-42}. Equation
(\ref{eq: evoluzione h}) can be recast in the form \begin{equation}
\ddot{h}+(3n+m)t^{-1}\frac{dh}{dt}+k^{2}a_{0}(\frac{t_{0}}{t})^{2n}h=0,\label{eq: evoluzione h 2}\end{equation}

whose general solution is \begin{equation}
h(t)=(\frac{t_{0}}{t})^{-\beta}[C_{1}J_{\alpha}(x)+C_{2}J_{-\alpha}(x)].\label{eq: sol ev h}\end{equation}

$J_{\alpha}$'s are Bessel functions and

\begin{equation}
\alpha=\frac{1-3n-m}{2(n-1)},\textrm{ $ $$ $$ $}\beta=\frac{1-3n-m}{2},\textrm{ $ $$ $$ $}x=\frac{kt^{1-n}}{1-n}\label{eq: alfabetx}\end{equation}

while $t_{0},$ $C_{1},$ and $C_{2}$ are constans related to the
specific values of $n$ and $m.$

The time units are in terms of the Hubble radius $H^{-1};$ $n=1/2$
is a radiation-like evolution; $n=2/3$ is a dust-like evolution,
$n=2$ labels power-law inflationary phases and $n=-5$ is a pole-like
inflation. From eq. (\ref{eq: m n}), a singular case is for $m=-2$
and $s=2.$ It is clear that the conformally invariant plane-wave
amplitude evolution of the tensor GW strictly depends on the background.

Let us now take into account the iusse of the production of relic
GWs contributing to the stochastic background. Several mechanisms
can be considered as cosmological populations of astrophysical sources
\cite{key-43}, vacuum fluctuations, phase transitions \cite{key-24}
and so on. In principle, we could seek for contributions due to every
high-energy physical process in the early phases of the Universe evolution.

It is important to distinguish processes coming from transitions like
inflation, where the Hubble flow emerges in the radiation dominated
phase and process, like the early star formation rates, where the
production takes place during the dust dominated era. In the first
case, stochastic GWs background is strictly related to the cosmological
model. This is the case we are considering here which is, furthermore,
also connected to the specific theory of gravity. In particular, one
can assume that the main contribution to the stochastic background
comes from the amplification of vacuum fluctuactons at the transition
between an inflationary phase and the radiation-dominated era. However,
in any inflationary model, we can assume that the relic GWs generated
as zero-point fluctuactions during the inflation undergoes adiabatically
damped oscillations ($\sim1/a$) until they reach the Hubble radius
$H^{-1}.$ This is the particle horizon for the growth of perturbations.
On the other hand, any other previous fluctuation is smoothed away
by the inflationary expansion. The GWs freeze out for $a/k\gg H^{-1}$
and re-enter the $H^{-1}$ radius after the reheating in the Friedman
era \cite{key-21,key-25,key-26,key-27,key-33}. The re-enter in the
radiation dominated or in the dust-dominated era depends on the scale
of the GW. After the re-enter, GWs can be detected by their Sachs-Wolfe
effect on the temperature anisotropy $\frac{\bigtriangleup T}{T}$
at the decoupling \cite{key-30}. When $\Phi$ acts as the inflaton
\cite{key-21,key-44} we have $\frac{d\Phi}{dt}\ll H$ during the
inflation. Considering also the conformal time $d\eta=dt/a$, eq.
(\ref{eq: evoluzione h}) reads\begin{equation}
h''+2\frac{\gamma'}{\gamma}h'+k^{2}h=0,\label{eq: evoluzione h 3}\end{equation}

where $\gamma=ae^{\Phi}$and derivation is with respect to $\eta.$
Inflation means that $a(t)=a_{0}\exp(Ht)$ and then $\eta=\int dt/a=1/(aH)$
and $\frac{\gamma'}{\gamma}=-\frac{1}{\eta}.$ The exact solution
of (\ref{eq: evoluzione h 3}) is \begin{equation}
h(\eta)=k{}^{-3/2}\sqrt{2/k}[C_{1}(\sin k\eta-\cos k\eta)+C_{2}(\sin k\eta+\cos k\eta)].\label{eq: sol ev h2}\end{equation}

Inside the $1/H$ radius it is $k\eta\gg1.$ Furthermore considering
the absence of gravitons in the initial vacuum state, we have only
negative-frequency modes and then the adiabatic behavior is \begin{equation}
h=k{}^{1/2}\sqrt{2/\pi}\frac{1}{aH}C\exp(-ik\eta).\label{eq: sol ev h3}\end{equation}

At the first horizon crossing ($aH=k$), the averaged amplitude $A_{h}=(k/2\pi)^{3/2}|h|$
of the perturbations is \begin{equation}
A_{h}=\frac{1}{2\pi{}^{2}}C\label{eq: Ah}\end{equation}

when the scale $a/k$ grows larger than the Hubble radius $1/H,$
the growing mode of evolution is constant, that it is frozen. This
situation corresponds to the limit $-k\eta\ll1$ in equation (\ref{eq: sol ev h2}).
Since $\Phi$ acts as the inflaton field, it is $\Phi\sim0$ at re-enter
(after the end of inflation). Then the amplitude $A_{h}$ of the wave
is preserved until the second horizon crossing after which it can
be observed, in principle, as an anisotropy perturbation of the CBR.
It can be shown that $\frac{\bigtriangleup T}{T}\lesssim A_{h}$ is
an upper limit to $A_{h}$ since other effects can contribute to the
background anisotropy \cite{key-45}. From this consideration, it
is clear that the only relevant quantity is the initial amplitude
$C$ in equation (\ref{eq: sol ev h3}) which is conserved until the
re-enter. Such an amplitude directly depends on the fundamental mechanism
generating perturbations. Inflation gives rise to processes capable
of producing perturbations as zero-point energy fluctations. Such
a mechanism depends on the adopted theory of gravitation and then
$\frac{\bigtriangleup T}{T}$ could constitute a further constraint
to select a suitable f(R)-theory. Considering a single graviton in
the form of a monocromatic wave, its zero-point amplitude is derived
through the commutation relations \begin{equation}
[h(t,x),\pi_{h}(t,y)]=i\delta^{3}(x-y)\label{eq: commutare}\end{equation}

calculated at a fixed time $t,$ where the amplitude $h$ is the field
and $\pi_{h}$ is the conjugate momentum operator. Writing the lagrangian
for $h$ 

\begin{equation}
\widetilde{L}=\frac{1}{2}\sqrt{\widetilde{g}}\widetilde{g}^{\alpha\beta}h_{;\alpha}h_{;\beta}\label{eq: lagrange}\end{equation}

in the conformal FRW metric $\widetilde{g}_{\alpha\beta}$ ($h$ si
conformally invariant), we obtain

\begin{equation}
\pi_{h}=\frac{\partial\widetilde{L}}{\partial(dh/dt)}=e^{2\Phi}a^{3}\frac{dh}{dt}.\label{eq: pi h}\end{equation}

Equation (\ref{eq: commutare}) becomes\begin{equation}
[h(t,x),\frac{dh}{dt}(y,y)]=i\frac{\delta^{3}(x-y)}{e^{2\Phi}a^{3}}\label{eq: commutare 2}\end{equation}

and the fields $h$ and $\frac{dh}{dt}$ can be expanded in terms
of creation and annihilation operators

\begin{equation}
h(t,x)=\frac{1}{(2\pi)^{3/2}}\int d^{3}k[h(t)e^{-ikx}+h^{*}(t)e^{ikx}]\label{eq: crea}\end{equation}
\begin{equation}
\frac{dh}{dt}(t,x)=\frac{1}{(2\pi)^{3/2}}\int d^{3}k[\frac{dh}{dt}(t)e^{-ikx}+\frac{dh}{dt}^{*}(t)e^{ikx}].\label{eq: distruggi}\end{equation}

The commutation relations in conformal time are then \begin{equation}
[h(h'^{*}-h^{*}h']=i\frac{(2\pi)^{3}}{e^{2\Phi}a^{3}}.\label{eq: commutare 3}\end{equation}

Insering (\ref{eq: sol ev h3}) and (\ref{eq: Ah}), we obtain $C=\sqrt{2}\pi{}^{2}He^{-\Phi}$
where $H$ and $\Phi$ are calculated at the first horizon crossing
and then \begin{equation}
A_{h}=\frac{\sqrt{2}}{2}{}He^{-\Phi},\label{eq: Ah2}\end{equation}

which means that the amplitude of GWs produced during inflation directly
depends on the given f(R) theory being $\Phi=\frac{1}{2}\ln f'(R).$
Explicitly, it is \begin{equation}
A_{h}=\frac{H}{\sqrt{2f'(R)}}.\label{eq: Ah3}\end{equation}
 This result deserves some discussion and can be read in two ways.
From one side the amplitude of relic GWs produced during inflation
depends on the given theory of gravity that, if different from general
relativity, gives extra degrees of freedom which assume the role of
inflaton field in the cosmological dynamics \cite{key-44}. On the
other hand, the Sachs-Wolfe effect related to the CMBR temperatue
anisotropy could constitute a powerful tool to test the true theory
of gravity at early epochs, i.e. at very high redshift. This probe,
related with data a medium \cite{key-46} and low redshift \cite{key-47},
could strongly contribute 

\begin{enumerate}
\item to reconstruct cosmological dynamics at every scale;
\item to further test general relativity or to rule out it against alternative
theories;
\item to give constrains on the SBGWs, if f(R) theories ares independently
probed at other scales.
\end{enumerate}
Reasuming, in this section it has been shown that amplitudes of tensor
GWs are conformally invariant and their evolution depends ond the
cosmological SBGWs. Such a background is tuned by conformal scalar
field which is not present in the standard general relativity. Assuming
that primordial vacuum fluctuations produce a SBGWs, beside scalar
perturbations, kinematical distorsions and so on, the initial amplitude
of these ones is function of the f(R)-theory of gravity and then the
SBGWs can be, in a certain sense, {}``tuned'' by the theory. Vice
versa, data coming fro the Sachs-Wolfe effect could contribute to
select a suitable f(R)-theory which can be consistently matched with
other observations. However, further and accurate studies are needed
in order to test the relation between Sachs-Wolfe effect and f(R)
gravity. This goal could be achieved in the next future through the
forthcoming space (LISA) and ground based (Virgo, LIGO) interferometers.

\section{Conclusions}

The SBGWs has been analyzed with the auxilium of the WMAP data while,
in general, in previous works in the literature about the SBGWs, old
COBE data were used. After this, it has been shown how the SBGWs and
f(R) gravity can be related, showing, vice versa, that a revealed
SBGWs could be a powerly probe for a given theory of gravity. In this
way, it has also been shown that the conformal treatment of SBGWs
can be used to parametrize in a natural way f(R) theories.


\begin{thebibliography}{10}
\bibitem{key-1}Peebles PJE and Ratra B - Rev. Mod. Phys. \textbf{75} 8559 (2003)
\bibitem{key-2}Allemandi G, Capone M, Capozziello S and Francaviglia M - Gen. Rev.
Grav. 38 1 (2006)
\bibitem{key-3}Allemandi G, Francaviglia M, Ruggiero ML and Tartaglia A - Gen. Rel.
Grav. 37 11 (2005)
\bibitem{key-4}Capozziello S - Int. J. Mod. Phys. D \textbf{11} 483 (2002);
\bibitem{key-5}Capozziello S, Cardone VF and Troisi A - J. Cosmol. Astropart. Phys.
JCAP08001 (2006)
\bibitem{key-6}Capozziello S - \textit{Newtonian Limit of Extended Theories of Gravity}
in \textit{Quantum Gravity Research Trends} Ed. A. Reimer, pp. 227-276
Nova Science Publishers Inc., NY (2005) - \foreignlanguage{italian}{also
in arXiv:}gr-qc/0412088 (2004) 
\bibitem{key-7}Capozziello S and Troisi A - Phys. Rev. D \textbf{72} 044022 (2005) 
\bibitem{key-8}Will C M \textit{Theory and Experiments in Gravitational Physics},
Cambridge Univ. Press Cambridge (1993)
\bibitem{key-9}Acernese F et al. (the Virgo Collaboration) - Class. Quant. Grav.
23 8 S63-S69 (2006) 
\bibitem{key-10}Corda C - Astropart. Phys. \textbf{27,} No 6, 539-549 (2007)
\bibitem{key-11}Hild S (for the LIGO Scientific Collaboration) - Class. Quant. Grav.
\textbf{23} 19 S643-S651 (2006)
\bibitem{key-12}Willke B et al. - Class. Quant. Grav. \textbf{23} 8S207-S214 (2006) 
\bibitem{key-13}Sigg D (for the LIGO Scientific Collaboration) - www.ligo.org/pdf\_public/P050036.pdf
\bibitem{key-14}Abbott B et al. (the LIGO Scientific Collaboration) - Phys. Rev. D
72, 042002 (2005) 
\bibitem{key-15}Ando M and the TAMA Collaboration - Class. Quant. Grav. \textbf{19}
7 1615-1621 (2002)
\bibitem{key-16}Tatsumi D, Tsunesada Y and the TAMA Collaboration - Class. Quant.
Grav. \textbf{21} 5 S451-S456 (2004) 
\bibitem{key-17}Capozziello S and Corda C - Int. J. Mod. Phys. D \textbf{15} 1119
-1150 (2006); Corda C - \textit{Response of laser interferometers
to scalar gravitational waves}- talk in the \textit{Gravitational
Waves Data Analysis Workshop in the General Relativity Trimester of
the Institut Henri Poincare -} Paris 13-17 November 2006, on the web
in www.luth2.obspm.fr/IHP06/workshops/gwdata/corda.pdf
\bibitem{key-18}Corda C - J. Cosmol. Astropart. Phys. JCAP04009 (2007)
\bibitem{key-19}Corda C - Astropart. Phys. 28, 247-250 (2007)
\bibitem{key-20}Corda C - Mod. Phys. Lett. A 22, 16, 1167-1173 (2007)
\bibitem{key-21}Capozziello S, Corda C and De Laurentis MF - Mod. Phys. Lett. A 22,
15, 1097-1104 (2007)
\selectlanguage{italian}
\bibitem{key-22}C. L. Bennet and others - ApJS \textbf{148} 1 (2003)
\bibitem{key-23}D. N. Spergel and others - ApJS \textbf{148} 195 (2003)
\selectlanguage{english}
\bibitem{key-24}Maggiore M- Physics Reports \textbf{331}, 283-367 (2000)
\bibitem{key-25}B. Allen - Proceedings of the Les Houches School on Astrophysical
Sources of Gravitational Waves, eds. Jean-Alain Marck and Jean-Pierre
Lasota (Cambridge University Press, Cambridge, England 1998).
\selectlanguage{italian}
\bibitem{key-26}L. P. Grishchuk and others - Phys. Usp. 44 1-51 (2001)
\bibitem{key-27}L. P. Grishchuk and others - Usp. Fiz. Nauk 171 3 (2001)
\bibitem{key-28}Babusci D, Foffa F, Losurdo G, Maggiore M, Mattone G and Sturani R
- Virgo DAD - www.virgo.infn.it/Documents/DAD/stochastic background
\selectlanguage{english}
\bibitem{key-29}Corda C - \textit{Virgo Report:} VIRGO-NOTE-PIS 1390-237 (2003) \foreignlanguage{italian}{-
www.virgo.infn.it/Documents}
\bibitem{key-30}R. K. Sachs and A. M. Wolfe - \foreignlanguage{italian}{ApJ 147, 73
(1967)}
\selectlanguage{italian}
\bibitem{key-31}B. Novosyadlyj and S. Apunevych - proceedings of international confernce
{}``Astronomy in Ukraine - Past, Present, Future'' - Main Astronomical
Observatory (2004)
\selectlanguage{english}
\bibitem{key-32}J. P. Zibin, D. Scott and M. White \foreignlanguage{italian}{- arXiv:astro-ph/9904228}
\bibitem{key-33}B. Allen - Phys. Rev. D \foreignlanguage{italian}{\textbf{3}}\textbf{7},
2078 (1988)
\bibitem{key-34}B. Allen and J. P. Romano - Phys. Rev. D \foreignlanguage{italian}{\textbf{59}
102001 (1999)}
\bibitem{key-35}E. E. Flanagan - Phys. Rev. D \foreignlanguage{italian}{\textbf{48}
2389 (1993)}
\bibitem{key-36}K. G. Arun, B. R. Iver, B. S. Sathyaprakash, and P. A. Sundararajan
- Phys. Rev. D \foreignlanguage{italian}{\textbf{71} 084008 (2005)}
\bibitem{key-37}www.lisa.nasa.org; www.lisa-scienze.org 
\bibitem{key-38}S. Weimberg - \textit{Gravitation and Cosmology} (Wiley 1972)
\bibitem{key-39}Maggiore M and Nicolis A - \foreignlanguage{italian}{Phys.} Rev. \foreignlanguage{italian}{D
\textbf{62} 024004 (2000)} 
\bibitem{key-40}Tobar ME, Suzuki T and Kuroda K \foreignlanguage{italian}{Phys.} Rev.
\foreignlanguage{italian}{D 59 \textbf{}102002 (1999)}
\bibitem{key-41}Corda C - Mod. Phys. Lett. A 22, No. 23, 1727-1735 (2007)
\bibitem{key-42}Capozziello S, Cardone VF, Carloni S and Troisi A - Int. J. Mod. Phys.
D \textbf{12} 1969 (2003)
\bibitem{key-43}Chiba T, Smith TL and Erickcek L - astro-ph/0611867 (2006)
\bibitem{key-44}Starobinsky AA - Phys. Lett. B \textbf{91}, 99 (1980)
\bibitem{key-45}Starobinsky AA - Sov. Phys. JEPT Lett. B \foreignlanguage{italian}{\textbf{34}},
438 (1982)
\bibitem{key-46}Capozziello S, Cardone VF and Troisi A - Phys. Rev. D \foreignlanguage{italian}{\textbf{71}
08}43503 (2005)
\bibitem{key-47}Capozziello S, Cardone VF , Funaro M and Andreon S - Phys. Rev. D
\foreignlanguage{italian}{\textbf{70} 123501} (2004)
\end{thebibliography}
\end{document}